\documentclass[12pt]{article}
\usepackage{psfrag}
\usepackage{graphicx}
\usepackage{float}
\usepackage{rotating}
\usepackage{longtable}
\usepackage{supertabular}
\usepackage{amssymb}
\usepackage{mathrsfs}
\usepackage{amsmath}

\title{Finite size scaling and first order phase transition in a modified XY-model\\[8mm]
}
\author{Suman Sinha
\footnote{E-mail: ssinha@research.jdvu.ac.in}
 ~and
Soumen Kumar Roy
\footnote{Corresponding author. E-mail: skroy@phys.jdvu.ac.in,
Tel: +91 9331910161; fax: +91 33 24146584}\\
Department of Physics,\\
Jadavpur University, Kolkata - 700032, INDIA}
\date{  }

\begin{document}

\maketitle
\begin{abstract}
Monte Carlo simulation has been performed in a two-dimensional modified XY-model first proposed by
Domany $et. al$ [E. Domany, M. Schick and R. H. Swendsen, Phys. Rev. Lett. {\bf 52}, 1535 (1984)].
The cluster algorithm of Wolff has been used and multiple histogram reweighting
is performed. The first order scaling behavior of the quantities like specific heat, order
parameter susceptibility and free energy barrier are found to be obeyed accurately. While the
lowest order correlation function was found to decay to zero at long distance just above
the transition, the next higher order correlation function shows a non-zero plateau.
\end{abstract}
{\it keywords:} Monte Carlo, Wolff, Finite Size Scaling
\\
\\

\section{\bf Introduction}
More than two decades ago Domany $et. al$ \cite{ds4} proposed a generalization of the two-dimensional XY-
model where the shape of the usual $\tt cos\theta$ type potential could be modified with the help of a
single parameter. The two-dimensional spins located at the sites of a square lattice interact with the
nearest neighbors through a potential
\begin{equation}
V(\theta_{ij})=2\Big[1-\Big(\tt cos^2{\frac{\theta_{ij}}{2}}\Big)^{p^2}\Big]
\label{poten}
\end{equation}
where $\theta_{ij}$ is the angle between the spins and $p^2$ is a parameter used to alter the shape of the
potential. For $p^2=1$ the potential reproduces the conventional XY-model while for larger values of
$p^2$ the potential well becomes narrower. The conventional two-dimensional XY model does not possess any true long range
order which is ruled out by the Mermin Wagner theorem. However a continuous quasi-long-range-order-disorder
transition resulting from the unbinding of topological defects \cite{kt4,kos4} is known to occur in this
system and the order parameter correlation function is characterized by a slow algebraic decay instead
of the fast exponential decay observed in a disordered system and this is referred to as the
Kosterlitz-Thouless (KT) transition in literature. Domany $et. al$ \cite{ds4} performed Monte Carlo (MC) simulation and
observed that as the potential well gets narrower with the increase in the parameter $p^2$, the
continuous transition gets converted into a first order phase transition and for $p^2=50$ the transition
is very sharp as is manifested by a huge peak in the specific heat. This phenomenon is in apparent contradiction
with the prediction of the renormalization group theory according to which systems in the same universal
class (having same symmetry of the order parameter and same lattice dimensionality) should exhibit the same
type of phase transition with identical values of critical exponents.

The generalized XY-model of Eqn.(\ref{poten}) has been analyzed by a number of authors \cite{knops4,him14}
using the
renormalization approach of the Migdal-Kadanoff type. These investigators were of the opinion that the
transition in the generalized XY-model appears to be first order in nature because the MC simulation of
Domany $et. al$ \cite{ds4} and Himbergen \cite{him4} were carried out on relatively small lattices and for
large system sizes the usual KT transition is expected to occur. Nearly a decade later, Mila \cite{mila4}
using the same sort of renormalization group analysis arrived at a similar conclusion. Lastly, using the
same line of approach, Garel $et. al$ \cite{garel4} put forward a different type of interpretation of the
above mentioned RG analysis and were of the view that the transition is indeed first order.

Minnhagen \cite{min14,min24,min34} has carried out a detailed study of the behavior of the phase transition
exhibited by a 2-D Coulomb gas, which very well describes the characteristics of a 2-D system consisting
of vortex-antivortex pairs. It was demonstrated that the KT behavior is obtainable in a 2-D Coulomb gas
only at low particle densities. For higher particle densities the charge unbinding transition was shown
to be first order. Also a new gas-liquid like critical point was found in the 2-D Coulomb gas --- the first
order line in the temperature-particle density plane ends at a critical point. The KT transition line,
obtainable at lower densities was seen to join smoothly with the first order line at a temperature slightly
lower than the critical point. Jonsson, Minnhagen and Nylen \cite{jmn4} performed MC simulation in a 2-D
XY-model with a modified potential, which essentially is equivalent to that of Eqn.(\ref{poten}) and established a
new critical point. They determined the critical exponents for the system and interpreted the transition
to be of the vortex unbinding type.

van Enter and Shlosman \cite{es4} presented a rigorous proof that various SO(n)-invariant
n-vector models which have a deep and narrow potential well, would exhibit a first order transition. The
model represented by Eqn.(\ref{poten}) is a member of this general class of systems. These authors based their proof
on the so called method of reflection positivity, a technique borrowed from the field theory and used in
statistical mechanics. van Enter and Shlosman argued that in spite of the order parameter in 2-D n-vector
model being predicted to vanish by the Mermin-Wagner theorem, long range order prevails in the system via
higher order correlation functions. More recently, S. Ota and S. B. Ota \cite{ota4} have performed MC
simulation of the modified XY- model using microcanonical ensemble and have identified a first order
phase transition in the system.

The present article describes MC simulation of the 2-D modified XY-model where computations have been
performed on systems of reasonably large size and finite size scaling rules for first order phase transition
have been tested on the results of the simulation. The motivation is to resolve the question on the nature
of the phase transition in this model and the contradictions among the views put forward by different
investigators for the last quarter of a century as has been summed up above.
Our observation is that the transition is indeed first order for a large value of the parameter
$p^2$ (we have used $p^2=50$) as all finite size scaling rules are nicely obeyed. We however have made no
attempt to investigate the existence of the critical point in this model or to determine the critical
exponents as has been done by Jonsson $et. al$ \cite{jmn4} in relatively small systems. Among other observables
we have computed the spin-spin angular correlation functions of different orders. We observe that while the
lowest order correlation function decays to zero, the next higher order correlation function has a finite
plateau which is in accordance with statement of van Enter and Sholsman \cite{es4}.

Another interesting aspect of our work is the application of the Wolff cluster algorithm \cite{wlf4}
to simulate
the model. It has been pointed out by a numbers of workers \cite{ds4,ssskr4} that the two-dimensional model is
difficult to simulate using the conventional single spin flip Metropolis algorithm \cite{metro4}.
To increase the reliability of the results we have used the multiple histogram reweighting, due to Ferrenberg and Swendsen
\cite{fs4} along with the Lee and Kosterlitz's method \cite{lk4} of finite size scaling for a first order phase
transition.

\section{\bf The definition of the thermodynamic quantities related to the model}

The Monte-Carlo simulation was carried out on a square lattice of dimension $L \times L$ with the
two-dimensional spins located at each site and interacting with the nearest neighbors via the
Hamiltonian
\begin{equation}
\displaystyle H=\sum_{{\langle}ij{\rangle}}2\Big[1-\Big(\tt cos^2{\frac{\theta_{ij}}{2}}\Big)^{p^2}\Big]
\end{equation}
The specific heat at a dimensionless temperature $T$ is related to the energy fluctuation
\begin{equation}
C_v={\frac{\displaystyle\left({\langle}H^2{\rangle}-{\langle}H{\rangle}^2\right)}{N\displaystyle T^2}}
\end{equation}
where $N$ is the number of spins.
The conventional long range order parameter is given by
\begin{equation}
{\langle}P_1{\rangle}={\langle}\tt cos \theta {\rangle}
\end{equation}
where $\theta$ is the angle that a spin makes with the preferred direction of orientation and the average
is over the entire sample. The next higher rank order parameter is defined as
\begin{equation}
{\langle}P_2{\rangle}={\frac{1}{2}{\langle}3\tt cos^2 \theta-1 {\rangle}}
\end{equation}
The order parameter susceptibility is defined in terms of the fluctuations of the order parameter
${\langle}P_1{\rangle}$
\begin{equation}
\chi={\frac{\displaystyle\left({\langle}P_1^2{\rangle}-{\langle}P_1{\rangle}^2\right)}{\displaystyle T^2}}
\end{equation}
The first rank pair correlation coefficient is defined as
\begin{equation}
G_1(r)={{\langle}\left(\tt cos \theta_{ij}\right){\rangle}}_r
\end{equation}
where $i$ and $j$ are two spins separated by a distance $r$.
The second rank pair correlation coefficient is defined as
\begin{equation}
G_2(r)={{\langle}P_2\left(\tt cos \theta_{ij}\right){\rangle}}_r
\end{equation}

\section{\bf The computational details}
In this section we briefly describe the Wolff cluster algorithm, the Ferrenberg-Swendsen multiple
histogram reweighting technique and the Lee-Kosterlitz finite size scaling for first order phase
transition.
The Monte-Carlo(MC) simulations were performed on square lattices of size $L^2$ for $L=16$, $32$, $64$,
 $96$, $128$, $160$ and $192$. We have used Wolff's cluster flip algorithm, the essential steps of which
are as follows.

\noindent (1) A random unit vector $\vec r$ is taken and a spin flip $\vec \sigma_x \rightarrow
\vec \sigma_x^{\prime}$ is defined as
\begin{equation}
\vec \sigma_x^{\prime}=\vec \sigma_x - 2\left(\vec \sigma_x, \vec r\right)\vec r
\end{equation}

\noindent (2) Bonds $\left(x, y\right)$ of the lattice are activated with a probability
\begin{equation}
P\left(x, y\right)=1-{\tt exp}\Big({\tt min} \lbrace 0, \beta S_7 \rbrace \Big)
\end{equation}
~where,

\begin{eqnarray}
&S_7&=S_6-S_5 \nonumber\\
&S_6&=2\left( \frac{1+S_1}{2}\right)^{p^2} \nonumber\\
&S_5&=2(S_4)^{p^2} \nonumber\\
&S_4&=\frac{\left(1+\left(S_1-2S_2S_3\right)\right)}{2} \nonumber\\
&S_3&=\left(\vec \sigma_y, \vec r\right)\nonumber\\
&S_2&=\left(\vec \sigma'_x, \vec r\right)\nonumber\\
&S_1&=\left(\vec \sigma'_x, \vec \sigma_y \right)\nonumber
\end{eqnarray}
~ This process leads to the formation of a cluster on the lattice.

\noindent (3) All spins in a cluster are now flipped according to $\vec \sigma_x \rightarrow
\vec \sigma_x^{\prime}$

We have calculated the thermodynamic quantities using multiple histogram reweighting technique of Ferrenberg
and Swendsen \cite {fs4}, which is briefly described below. The partition function of the system is
given by
\begin{equation}
Z\left(\beta\right)=\sum_E \rho\left(E\right){\tt exp}\left[-\beta E\right]
\end{equation}
where $\rho(E)$ is the density of states, $\beta=1/T$ (the Boltzmann constant has been set equal
to unity) and $E$ is the energy of the system. In the histogram reweighting method, energy histograms are
generated at a number of temperatures $\beta_i$ with $i=1,2,....R$ and $N_i(E)$ is the histogram count
for the $i^{th}$ temperature. We denote by $n_i$ the total number of configurations generated in the
$i^{th}$ simulation, i.e, $\displaystyle n_i=\sum_E N_i(E)$. According to references \cite{fs4} and
\cite{nb4}, the best estimate of the density of states, obtained after histogram reweighting,
 is given by
\begin{equation}
\rho\left(E\right)=\frac{\displaystyle\sum_{i=1}^R g_i^{-1} N_i\left(E\right)}
{\displaystyle\sum_{j=1}^R n_j g_j^{-1} Z_j^{-1} {\tt exp} \left[-\beta_j E\right]}
\end{equation}
where, $g_i=1+2\tau_i$, $\tau_i$ being the auto-correlation time for energy at the $i^{th}$ temperature.
Substituting eqn.(13) in eqn.(12) gives us a self consistent equation for the partition function
at any temperature $\beta$:
\begin{equation}
Z(\beta)=\sum_E \frac{\displaystyle\sum_i g_i^{-1} N_i(E) {\tt exp}[-\beta E]}
{\displaystyle\sum_j g_j^{-1} n_j Z_j^{-1} {\tt exp}[-\beta_j E]}
\end{equation}

One can also carry out the computation in terms of the probability instead of the partition function.
The unnormalized probability for an energy $E$ in the $k^{th}$ simulation is given by
\begin{equation}
p_k(E)=\rho(E){\tt exp}[-\beta_k E]
\end{equation}
i,e.,
\begin{equation}
Z(\beta_k)=\sum_E p_k(E)
\end{equation}
The free energy at the temperature $\beta_k$ is
\begin{equation}
f_k=-\frac{1}{\beta_k} {\tt ln}Z(\beta_k)
\end{equation}
i.e.,
\begin{equation}
{\tt exp}(-\beta_k f_k)=\sum_E p_k(E)=Z(\beta_k)
\end{equation}
In place of eqn.(14) we get a self consistent equation for the reweighted probability,
\begin{equation}
p_k(E)=\frac{\displaystyle\sum_i g_i^{-1} N_i(E) {\tt exp}[-\beta_k E]}
{\displaystyle\sum_j g_j^{-1} n_j {\tt exp}[\beta_j (f_j-E)]}
\end{equation}
The best estimate of the energy or any other observable $Q$ is given by \cite{nb4}
\begin{equation}
{\langle}Q{\rangle}=\frac{1}{Z(\beta)}\sum_{i,s} \frac{g_i^{-1} Q_{is} {\tt exp}[-\beta E_{is}]}
{\displaystyle\sum_j n_j g_j^{-1} Z_j^{-1} {\tt exp} [-\beta_j E_{is}]}
\end{equation}
where $Q_{is}$ is the histogram count of the observable $Q$ in the state $s$ obtained during the
$i^{th}$ simulation and $E_{is}$ is the total energy of such a state. The factors $g_k$ now correspond
to the observable $Q$.
 Lee and Kosterlitz \cite{lk4} proposed a
convenient method for the determination of the order of the phase transition which can be applied to
 systems having linear dimension less than the correlation length. For a temperature driven
first order transition in a finite system of volume $L^d$ with periodic boundary condition one needs to compute
the histogram count of the energy distribution denoted by $N(E;\beta,L)$. The $p^2=50$ model has a
characteristic double peak structure for $N(E;\beta,L)$ in the neighborhood of the transition temperature. The
two peaks of $N$ at $E_1(L)$ and $E_2(L)$ corresponding respectively to the ordered and disordered phases are
separated by a minimum at $E_m(L)$. A free-energy-like quantity is defined as
\begin{equation}
A(E;\beta,L,\mathcal N)=-\ln N(E;\beta,L)
\end{equation}
where $\mathcal N$ is the number of configurations generated. The quantity $A(E;\beta,L,\mathcal N)$ differs
from the free energy $F(E;\beta,L)$ by a temperature and $\mathcal N$ dependent additive quantity. A bulk free
energy barrier can therefore be defined as
\begin {equation}
\Delta F(L)=A(E_m;\beta,L,\mathcal N)-A(E_1;\beta,L,\mathcal N)
\end{equation}
It may be noted that at the transition temperature, $A(E_1;\beta,L,\mathcal N)=A(E_2;\beta,L,\mathcal N)$
and $\Delta F$ is independent of $\mathcal N$. For a continuous transition $\Delta F(L)$ is independent of
$L$ and for a temperature driven first order transition it is an increasing function of $L$, even when $L$
is smaller than the correlation length, $\xi$, prevailing at the system at the transition temperature. If one
is in a region where $L$ is much greater than $\xi$, then $\Delta F$ obeys the scaling relation \cite{lk4}
\begin{equation}
\Delta F \sim L^{d-1}
\end{equation}
Clearly, the temperature at which the double-well structure of $A$ has two equally deep minima gives a
precise estimation of the transition temperature.

\section{\bf Results and discussion}
\begin{figure}[!h]
\begin{center}
\resizebox{120mm}{!}{\rotatebox{-90}{\includegraphics[scale=1.2]{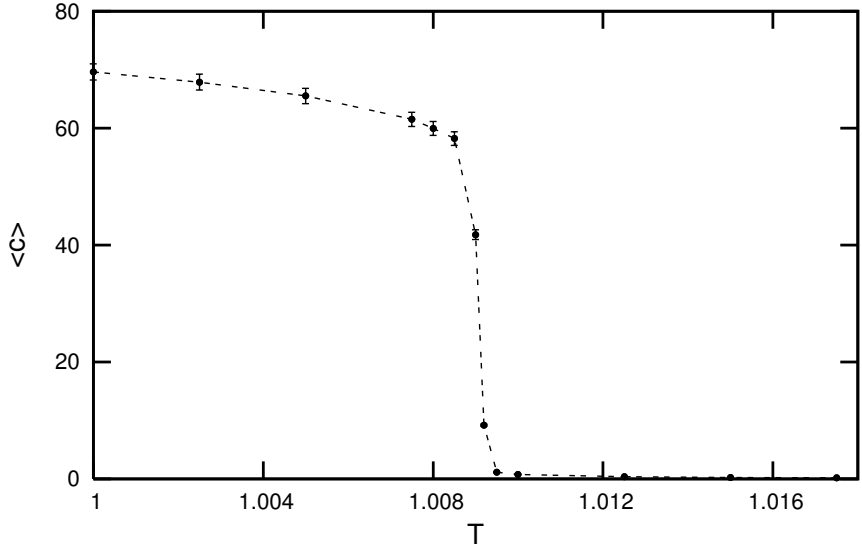}}}
\end{center}
\caption{The average cluster size ${\langle}c{\rangle}$ in percents of the lattice size for
$L=128$ obtained using the Wolff algorithm.}
\label{size4}
\end{figure}

\begin{figure}[!h]
\begin{center}
\resizebox{100mm}{!}{\rotatebox{-90}{\includegraphics[scale=1.2]{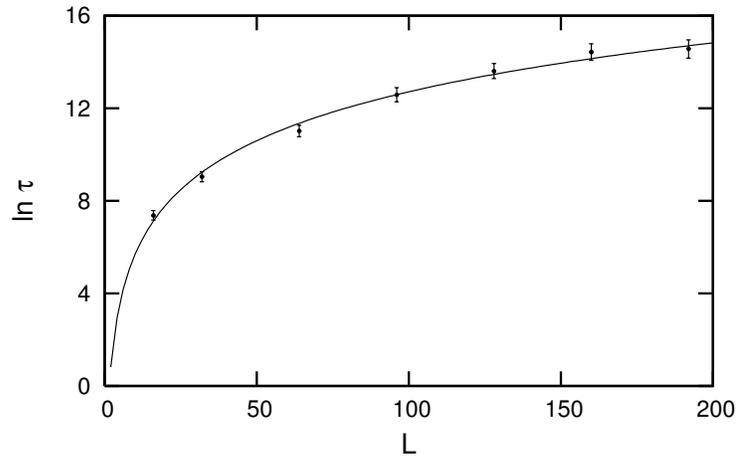}}}
\end{center}
\caption{Logarithmic plot of the peak value of the energy auto-correlation time against $L$.}
\label{tau4}
\end{figure}

Square lattices of linear dimension $L$ ranging from $16$ to $192$ were simulated and for each lattice simulations
were performed at $9$ to $13$ temperatures in the neighborhood of the transition to record the
histograms for energy. In Table~\ref{data}
we have depicted for each lattice size and temperature the technical quantities of interest. These
include the number of Wolff clusters generated $(n_c)$, the percentage of average cluster size
in units of lattice size for each temperature
${\langle}c{\rangle}$, the number of equivalent Monte Carlo sweeps (MCS) and the energy auto-correlation
time $\tau_e$. The number of configurations generated ranges from about $10^8$ to $10^9$.
 In Fig.\ref{size4}  we have plotted the percentage of average cluster size in units of the
lattice size ${\langle}c{\rangle}$ against temperature for $L=128$. It is clear that the average cluster size for a
given lattice,
decreases with increase in temperature and there is a sharp fall at the transition.
The maximum cluster size in units of the lattice size is about $84.4\%$ (for $L=16$) and is seen to decrease
with increase in the system size.
 The auto-correlation time, which was calculated by the method proposed by Madras and Sokal
\cite{ms4}, is seen to increase rapidly with the increase in lattice size and possesses a sharp maximum at the
transition temperature. The logarithm of the peak value of the energy auto-correlation time has been
plotted against $L$ in Fig.\ref{tau4} . We find empirically a scaling rule, $\ln \tau_e \sim L^\phi$ where
$\phi=3.05$.
\begin{table}[!h]
\caption{\footnotesize $T$ is the dimensionless temperature, L is the lattice size, $n_c$ is the number of Wolff clusters
, which varies from $1.1\times10^8$ to $10^9$ with lattice sizes, ${\langle}c{\rangle}$ is the average
cluster size as percent of the lattice size,
MCS is the number of equivalent Monte Carlo sweeps in units of $10^8$ and $\tau_e$
is the auto-correlation time for energy (in units of Wolff clusters).}
\begin{center}\
\resizebox{\textwidth}{!}{%
\begin{tabular}{|cccccccccccccc|}
\hline
& & & & & &${\bf L=192}$ & & & & & & & \\
$T$ &1.0025&1.0050&1.0075&1.0085&1.0087&1.0089&1.0090&1.0091&1.0093&1.0100&1.0112&1.0125& \\
$n_c$ &$10^9$&$10^9$&$10^9$&$10^9$&$10^9$&$10^9$&$10^9$&$10^9$&$10^9$&$10^9$&$10^9$&$10^9$& \\
${\langle}c{\rangle}$&67&64&60&56&56&50&5&2&0.59&0.34&0.21&0.16& \\
MCS &6.71&6.46&6.04&5.67&5.60&5.06&0.586&0.233&0.059&0.034&0.021&0.016&\\
$\tau_e$ &239&349&808&2522&4342&211246&1802195&1194159&82892&57193&55714&53209& \\
\hline
& & & & &${\bf L=160}$ & & & & & & & & \\
$T$ &1.0025&1.0050&1.0075&1.0087&1.0090&1.0093&1.0100&1.0112&1.0125& & & & \\
$n_c$ &$10^9$&$10^9$&$10^9$&$10^9$&$10^9$&$10^9$&$10^9$&$10^9$&$10^9$& & & &\\
${\langle}c{\rangle}$&67&65&60&56&18&0.91&0.49&0.31&0.23& & & &\\
MCS &6.74&6.50&6.08&5.60&1.81&0.09&0.049&0.031&0.023&&&&\\
$\tau_e$ &239&354&788&3873&1851577&54589&44409&37398&39685& & & &\\
\hline
& & & & & & &${\bf L=128}$ & & & & & & \\
$T$ &1.0000&1.0025&1.0050&1.0075&1.0080&1.0085&1.0090&1.0092&1.0095&1.0100&1.0125&1.0150&1.0175 \\
$n_c$ &$10^9$&$10^9$&$10^9$&$10^9$&$10^9$&$10^9$&$10^9$&$10^9$&$10^9$&$10^9$&$10^9$&$10^9$&$10^9$\\
${\langle}c{\rangle}$&69&67&65&61&59&58&41&9&1&0.76&0.36&0.25&0.19\\
MCS &6.96&6.78&6.55&6.14&5.99&5.82&4.17&0.92&0.11&0.076&0.036&0.025&0.019\\
$\tau_e$ &193&242&344&780&1111&1906&838017&814172&33708&27202&24802&21925&20478\\
\hline
& & & & &${\bf L=96}$ & & & & & & & & \\
$T$ &1.0000&1.0025&1.0050&1.0075&1.0080&1.0085&1.0090&1.0093&1.0100& & & & \\
$n_c$ &$9\times10^8$&$9\times10^8$&$9\times10^8$&$9\times10^8$&$9\times10^8$&$9\times10^8$&$9\times10^8$&$9\times10^8$&$9\times10^8$& & & &\\
${\langle}c{\rangle}$&70&67&65&62&60&58&47&20&1& & & &\\
MCS &6.30&6.14&5.93&5.58&5.45&5.23&4.25&1.86&1.33&&&&\\
$\tau_e$ &185&233&335&658&3720&49156&291844&139640&20637& & & &\\
\hline
& & & & & & ${\bf L=64}$ & & & & & & & \\
$T$ &1.0000&1.0025&1.0050&1.0075&1.0081&1.0087&1.0093&1.0100&1.0125&1.0150&1.0175&1.0200& \\
$n_c$ &$8\times10^8$&$8\times10^8$&$8\times10^8$&$8\times10^8$&$8\times10^8$&$8\times10^8$&$8\times10^8$&$8\times10^8$&$8\times10^8$&$8\times10^8$&$8\times10^8$&$8\times10^8$& \\
${\langle}c{\rangle}$&70&69&66&63&61&56&43&15&1&1&0.78&0.63& \\
MCS &5.66&5.52&5.34&5.04&4.92&4.55&3.46&1.27&0.11&0.08&0.06&0.05&\\
$\tau_e$ &181&229&319&977&2451&37018&61038&54839&6548&5755&5714&5519& \\
\hline
& & & & & &${\bf L=32}$ & & & & & & & \\
$T$ &0.9900&1.0000&1.0050&1.0075&1.0100&1.0112&1.0125&1.0137&1.0150&1.0200&1.0300&1.0400& \\
$n_c$ &$1.7\times10^8$&$1.7\times10^8$&$1.7\times10^8$&$1.7\times10^8$&$1.7\times10^8$&$1.7\times10^8$&$1.7\times10^8$&$1.7\times10^8$&$1.7\times10^8$&$1.7\times10^8$&$1.7\times10^8$&$1.7\times10^8$& \\
${\langle}c{\rangle}$&76&72&68&65&52&39&23&11&7&2&1&1& \\
MCS &1.29&1.23&1.16&1.10&0.88&0.67&0.39&0.19&0.11&0.04&0.02&0.019&\\
$\tau_e$ &110&170&561&2149&7488&8420&7941&6152&4070&1556&1355&1290& \\
\hline
& & & & & & &${\bf L=16}$ & & & & & & \\
$T$ &0.9500&0.9750&1.0000&1.0062&1.0125&1.0188&1.0219&1.0250&1.0312&1.0375&1.0500&1.0800&1.1000 \\
$n_c$ &$1.1\times10^8$&$1.1\times10^8$&$1.1\times10^8$&$1.1\times10^8$&$1.1\times10^8$&$1.1\times10^8$&$1.1\times10^8$&$1.1\times10^8$&$1.1\times10^8$&$1.1\times10^8$&$1.1\times10^8$&$1.1\times10^8$&$1.1\times10^8$ \\
${\langle}c{\rangle}$&84&80&74&66&57&38&25&17&8&5&3&2&1\\
MCS &0.93&0.889&0.816&0.734&0.631&0.425&0.275&0.197&0.094&0.060&0.042&0.025&0.021\\
$\tau_e$ &73&82&269&657&1226&1593&1560&1259&736&507&332&319&281\\
\hline
\end{tabular}}
\end{center}
\label{data}
\end{table}
\noindent
The behavior of the order parameter correlation time $\tau_o$ is found to be similar in nature
as that of $\tau_e$.
\begin{figure}[!h]
\begin{center}
\resizebox{100mm}{!}{\rotatebox{-90}{\includegraphics[scale=1.2]{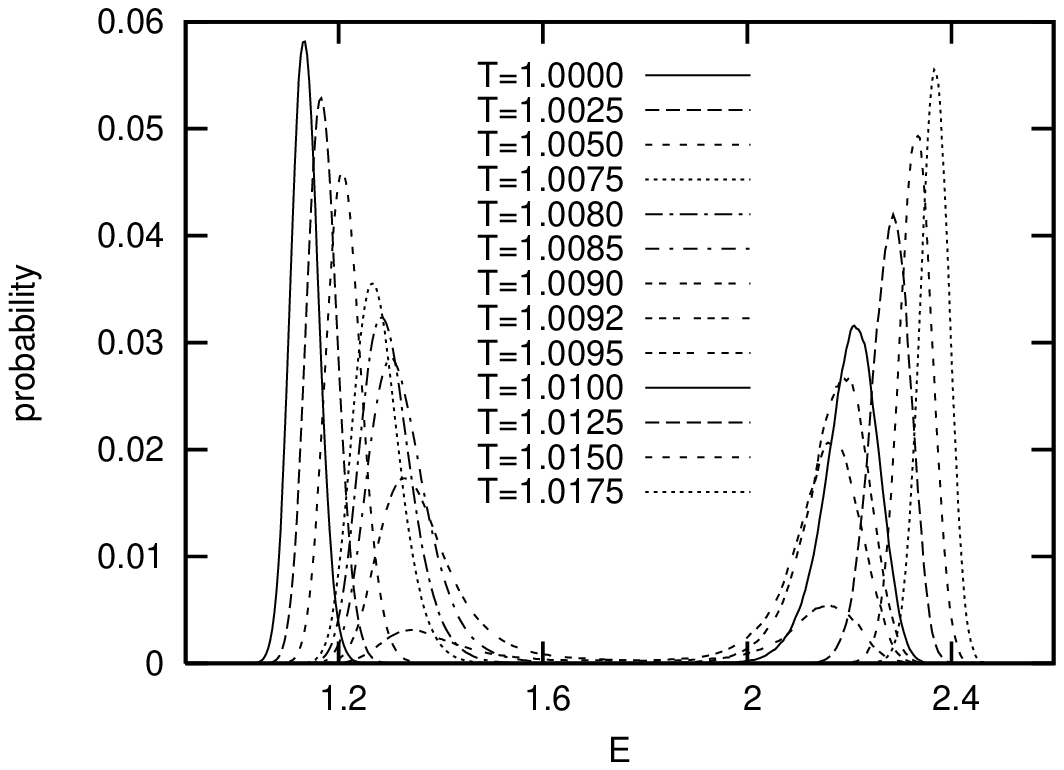}}}
\end{center}
\caption{The histograms for $E$, the average energy per particle generated for the
$128\times128$ lattice for the $p^2=50$ model at the $13$ temperatures indicated.}
\label{enghist4}
\end{figure}

The energy histograms obtained for $L=128$ are shown in Fig.\ref{enghist4}.
 For this lattice, simulations were performed at $13$ temperatures ranging
from $1.0000$ to $1.0175$. This temperature range is rather small and were chosen to bracket the
transition temperature. This diagram shows that there is an energy range where
almost no sampling takes place for any temperature and there are dual peaked histograms at a number
of temperatures where sampling takes place with one peak in the ordered phase and the other in the
disordered phase. The existence of these dual peaked histograms is a signature of a first order phase
transition where two phases can coexist at a given temperature.
\begin{figure}[!h]
\begin{center}
\resizebox{100mm}{!}{\rotatebox{-90}{\includegraphics[scale=1.2]{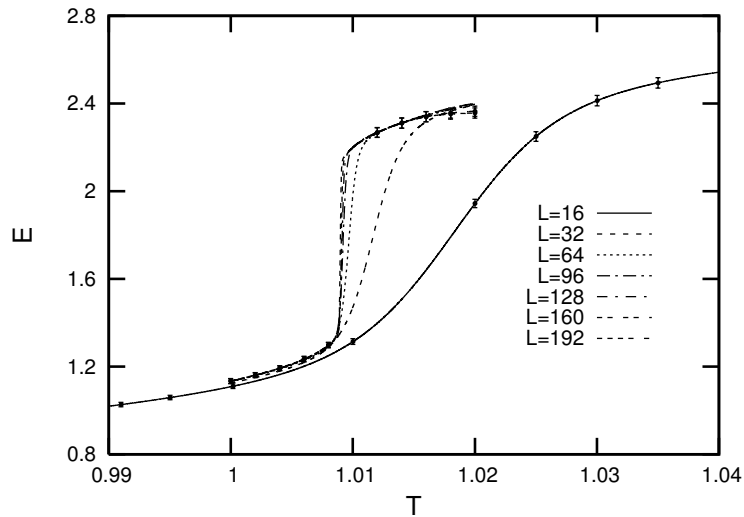}}}
\end{center}
\caption{The average energy per particle $E$ plotted against dimensionless temperature $T$ for different
lattice sizes.
 For clarity error bars are shown only for $L=16$,
$160$ and $192$.}
\label{eng4}
\end{figure}

The error in estimating the reweighted probability $p_k(Q)$ from the raw histograms is given by
\begin{equation}
\delta p_k\left(Q\right)=\frac{1}{{\left[\displaystyle\sum_{n=1}^R g_n^{-1}(q)N_n(q)\right]}^{1/2}}p_k\left(Q\right)
\end{equation}
and this can be estimated directly from the histogram counts. The percentage error in the
reweighted probability for energy in the lattice $L=192$ is about $0.74\%$
 where the raw histograms have peaks in the ordered phase. In the intermediate energy range
where little sampling takes place for any choice of temperature the error is evidently large and this cannot
be significantly reduced by any realistic effort.
\begin{figure}[!h]
\begin{center}
\resizebox{100mm}{!}{\rotatebox{-90}{\includegraphics[scale=1.2]{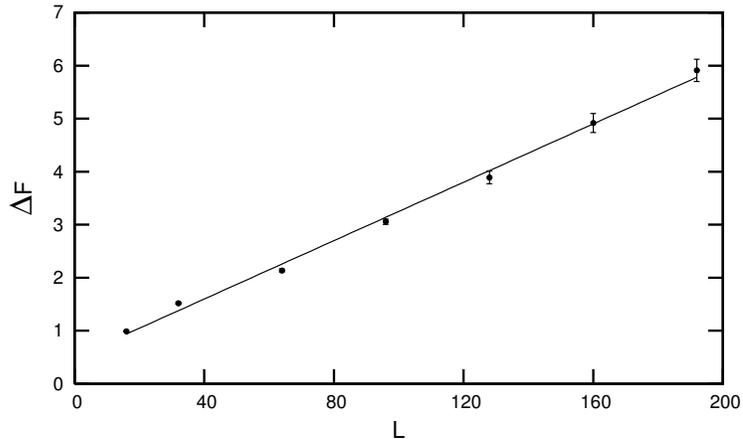}}}
\end{center}
\caption{The free-energy barrier height $\Delta F$ plotted against lattice size $L$ with the linear fit
represented by straight line.}
\label{fssfe4}
\end{figure}

Fig.\ref{eng4} shows the temperature variation of the energy for a number of lattices, as is obtained
by applying
histogram reweighting technique. From the energy histograms, we have calculated the free energy like quantity $A$, defined as
$A(E;\beta,L,\mathcal N)=-\ln N(E;\beta,L)$ where $N(E;\beta,L)$ is the histogram count of
the energy distribution. The free energy barrier $\Delta F(L)$ was evaluated and in Fig.\ref{fssfe4}
we have plotted $\Delta F$ against $L$ where a good linear fit has been obtained. This is a direct
verification of the scaling rule $\Delta F \sim L^{d-1}$ of Lee and Kosterlitz \cite{lk4} since the lattice
dimensionality $d=2$ in this model. We further note that the scaling relation is well obeyed down to
$L=16$ which happens to be of the order of the correlation length, $\xi$ for the system, as one can
estimate from the relation $\Delta F\left(\xi\right) \simeq 1$ \cite {lk4}.
\begin{figure}[!h]
\begin{center}
\resizebox{100mm}{!}{\rotatebox{-90}{\includegraphics[scale=1.2]{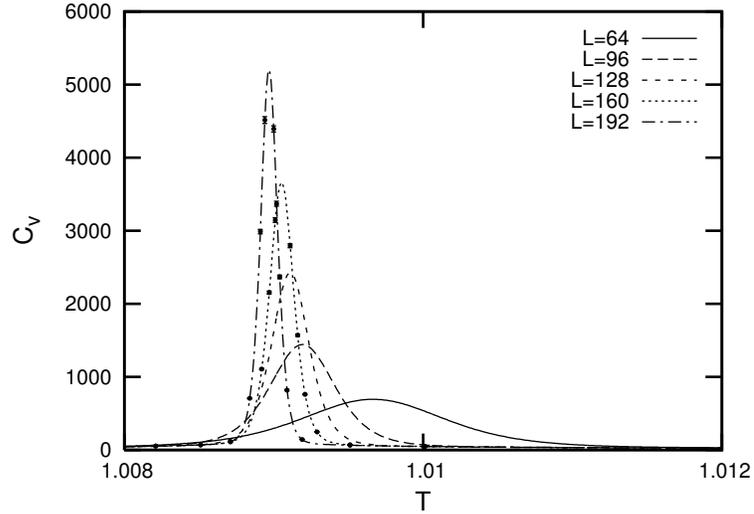}}}
\end{center}
\caption{The specific heat $C_v$ plotted against temperature $T$ for different lattice sizes.
 For clarity only the above lattice sizes are shown and the error bars have been indicated for two lattice
size.}
\label{cv4}
\end{figure}
\begin{figure}[!h]
\begin{center}
\resizebox{100mm}{!}{\rotatebox{-90}{\includegraphics[scale=1.2]{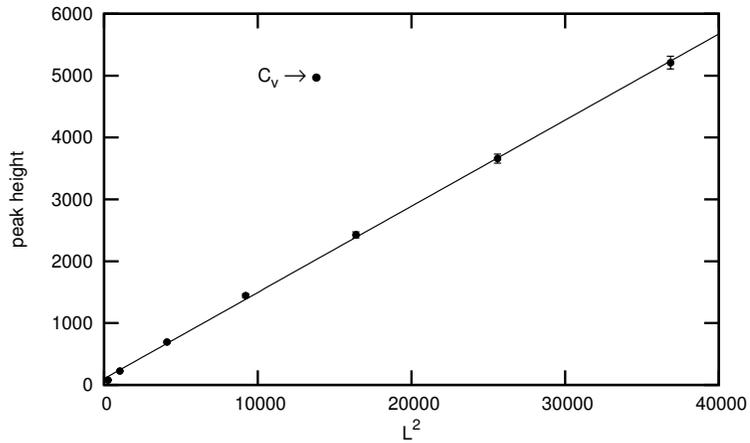}}}
\end{center}
\caption{The peak heights of $C_v$ plotted against $L^2$ with the linear fit represented by the
straight line. The error bars for most points are smaller than the dimensions of the symbols
used for plotting.}
\label{fsscv4}
\end{figure}

The specific heat $C_v$ was obtained from the energy fluctuation and Fig.\ref{cv4} shows its temperature variation.
It is evident that the peak height of $C_v$ grows rapidly at the transition. From Fig.\ref{fsscv4}, where the maxima of
$C_v$ are plotted it is clear that the standard scaling rules $C_v \sim L^d$ for first order transition
\cite{binder4}
are accurately obeyed in this model.
\begin{figure}[!h]
\begin{center}
\resizebox{100mm}{!}{\rotatebox{-90}{\includegraphics[scale=1.2]{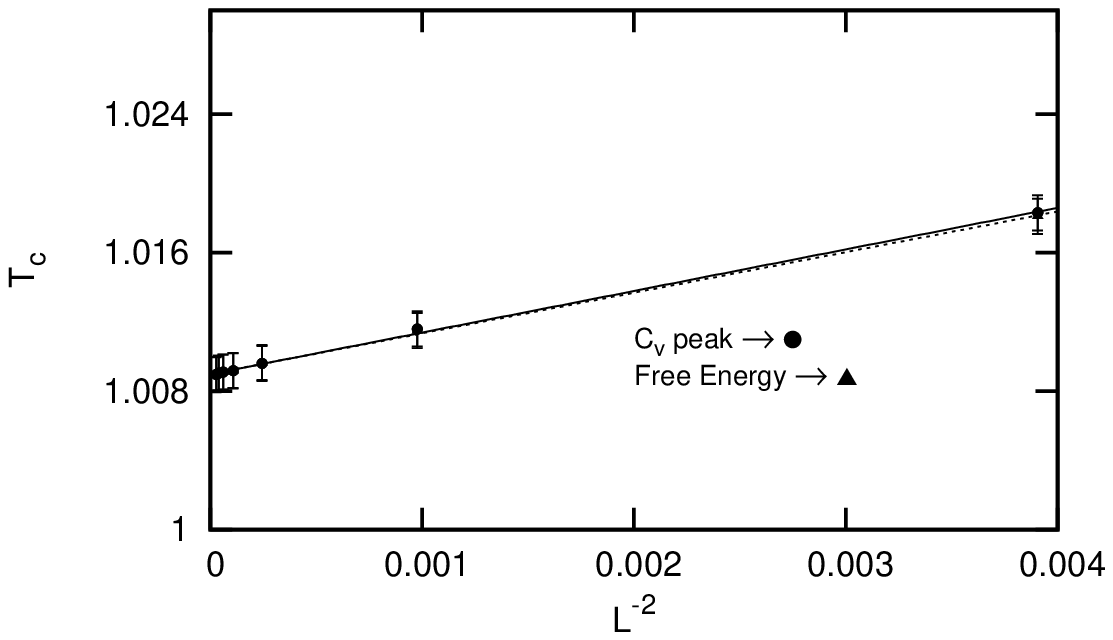}}}
\end{center}
\caption{The transition temperature $T_c$ obtained from (a) specific heat peak position and (b) fine tuning
of free energy curve plotted against $L^{-2}$ along with the
respective linear fits. The intercept on the Y-axis is $1.00897\pm6\times10^{-5}$.}
\label{fsstemp4}
\end{figure}

We have also tested the the finite size scaling relation
\begin{equation}
T_c\left(L\right)-T_c(\infty) \sim L^{-d}
\label{scltemp}
\end{equation}
which is valid for a first order phase transition \cite {binder4}. $T_c(\infty)$ represents the thermodynamic
limit of the transition temperature $T_c$. We have estimated the transition temperature in  two ways
--- $T_c^{C_v}$ is the estimate of $T_c$ obtained from the peak position of the specific heat $C_v$ and $T_c^F$
represents the transition temperature obtained from the fine tuning of the free energy vs energy curve to obtain
two equally deep minima. In Fig.\ref{fsstemp4} the
transition temperatures thus obtained have been plotted against $L^{-2}$. It is seen that the linear fits are
good within statistical errors and the thermodynamic limit of the transition temperature is
$1.00897\pm 6\times 10^{-5}$, within which the two linear fits are seen to converge.
\begin{figure}[!h]
\begin{center}
\resizebox{100mm}{!}{\rotatebox{-90}{\includegraphics[scale=1.2]{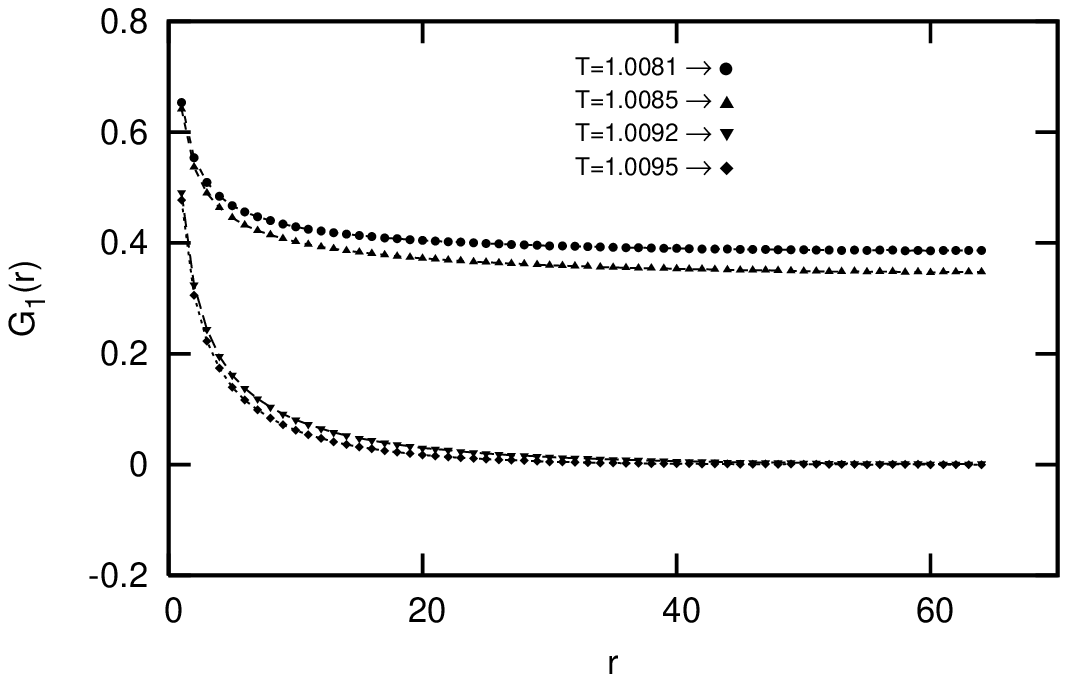}}}
\end{center}
\caption{The plots of the pair correlation function $G_1(r)$ against $r$ for the $128\times128$ lattice
for the temperatures indicated. The curves are plotted for $r$ ranging up to $L/2$.}
\label{angcor14}
\end{figure}
\begin{figure}[!h]
\begin{center}
\resizebox{100mm}{!}{\rotatebox{-90}{\includegraphics[scale=1.2]{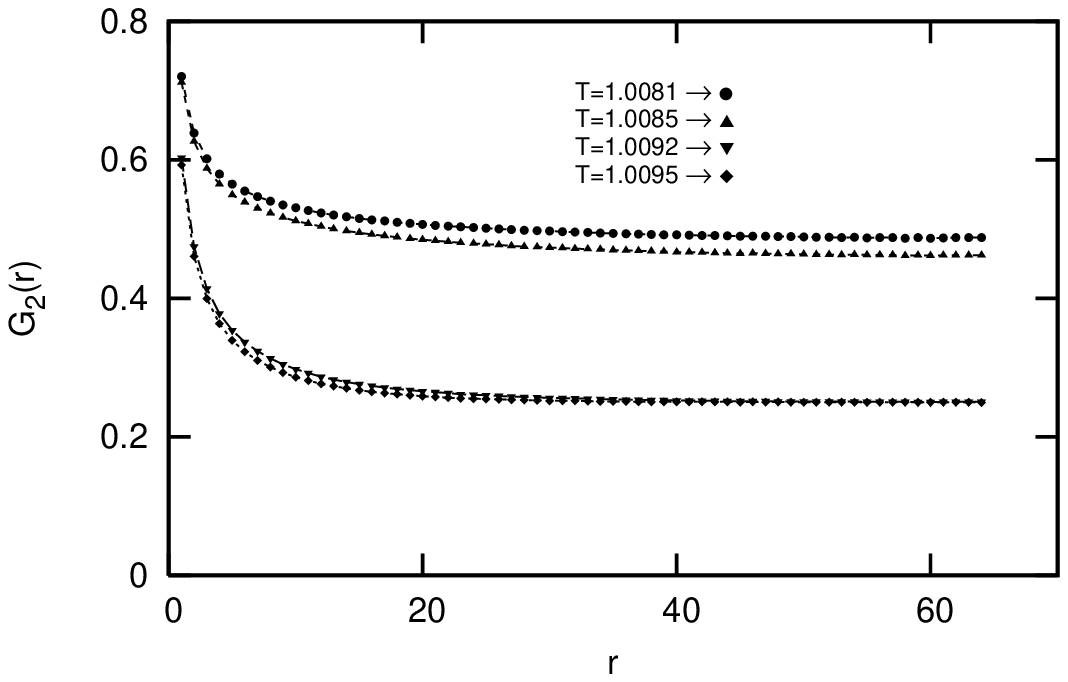}}}
\end{center}
\caption{The plots of the pair correlation function $G_2(r)$ against $r$ for the $128\times128$ lattice
for the temperatures indicated. The curves are plotted for $r$ ranging up to $L/2$.}
\label{angcor24}
\end{figure}

The pair correlation functions $G_1(r)$ and $G_2(r)$ were calculated for temperatures $T=1.0081$,
$1.0085$, $1.0092$ and $1.0095$ for $L=128$ and are shown in Fig.\ref{angcor14} and and \ref{angcor24}. The first two of these
temperatures are less than the transition temperature for this lattice while the other two temperatures are in the disordered
phase. The curves have been fitted to a power law $G_i(r)=a_ir^{-b_i}+f_i$ for $i=1$ and $2$. It may be noted that the parameter
$f$ is the asymptotic value of the pair correlation function. We observe that while the first rank correlation function
$G_1(r)$ decays to zero at the two higher temperatures ($f_1=0$), this is not the case for the higher rank
correlation function $G_2(r)$ ($f_2 \sim 0.22$). In other words, while the lowest rank correlation among the spins
vanishes just above transition, the next higher rank correlation continues to persist.

\section{\bf Conclusions}

The simulations in the two dimensional modified XY-model presented in this communication show that
all the first order finite size scaling rules are obeyed. Computation has been performed in
system size up to $192 \times 192$ which may be considered to be reasonably large for the purpose of
arriving at a conclusion regarding the behavior of the model. We are inclined to conclude that the
model exhibits a first order phase transition. This is in agreement with the views of some of the
earlier investigators including Domany $et. al$ \cite{ds4} and van Enter and Shlosman \cite{es4}. The
existence of a quasi-long-range-order-disorder transition observed in the 2-D XY-model is known to be
due to vortex-antivortex unbinding (KT transition). In absence of the role played by the vortices, one
would not observe any order-disorder transition in the XY-model in accordance with the Mermin-Wagner
theorem. In the class of models we have investigated the role played by the vortices changes qualitatively
with change in $p^2$ (which increases the non-linearity of the potential well) as
has been seen in the early work of Himbergen \cite{him4}. Also we have seen that the number of vortex
pairs grows rapidly with the increase in $p^2$ \cite{ssskr34}. Qualitatively, one may therefore think
that the modified XY-model for large values of $p^2$, behaves like
a dense defect system and gives rise to a first order phase transition
as has been predicted by Minnahagen \cite{min14, min24, min34}.

A similar change in the nature of phase transition has been observed to occur in a two-dimensional
Lebwohl-Lasher model and a modified version of it \cite{apskr4}. The potentials in these two models
are $-P_2({\tt cos}\theta)$ and $-P_4({\tt cos}\theta)$ respectively, the latter having a greater amount
of non-linearity. Although both models are in the same universality class, it was observed
that while the $-P_2({\tt cos}\theta)$ potential leads to a continuous transition, the modified model with
$-P_4({\tt cos}\theta)$ potential exhibits a strong first order phase transition. It has also been noticed
that the suppression of the defects in these models leads to a total disappearance of the phase transitions
\cite{sdskr4}.

We mention another point before ending this section. This is the performance of Wolff cluster algorithm
which turned out to be very convenient to simulate the model. Conventional algorithms, as we have seen,
does not work well in this model. Our earlier attempt \cite{ssskr4} using the recently developed Wang-Landau (WL)
algorithm \cite{wl4} which directly determines the density of states of a system is also not a good choice for
simulating this model. The main problem while using the WL algorithm is that configurations near the
minimum energy take a very long time to be sampled during the random walk and it becomes impractical to
simulate continuous models of even moderate size because of the huge CPU time that becomes necessary.
Among other things, a great virtue of the Wolff algorithm is that it does not
contain any adjustable parameter even while simulating a continuous model.

Besides using the Wolff algorithm for the simulation we have used the Ferrenberg-Swendsen multiple
histogram reweighting technique and the finite size scaling rules of Lee and Kosterlitz. We conclude
by noting that a combination of these computational tools till now provides a very efficient and
accurate method of analyzing results obtained in an unknown system.

\section{\bf Acknowledgment}

        The authors acknowledge the award of a CSIR (India) research grant 03(1071)/06/EMR-II which enabled us
to acquire four IBM X 226 servers, with which the work was done. One of us (SS) acknowledges the award
of a fellowship from the same project.

\end{document}